\documentclass[twoside,a4paper,10pt,twocolumn]{proceedings}
\usepackage{graphicx}
\usepackage{natbib}

\newcommand{\farcs}{\mbox{\ensuremath{.\!\!^{\prime\prime}}}}
\def\1{\c{c}}
\def\2{\c{C}}
\def\3{\.{I}}
\def\4{\"{a}}
\def\5{{\i}}
\def\6{$\beta$}
\def\7{\"{o}}
\def\8{\"{O}}
\def\9{\c{s}}
\def\0{\c{S}}
\def\*{\"{u}}
\def\;{\u{g}}
\def\:{\u{G}}

\begin{document}

\pagenumbering{arabic}
\pagestyle{myheadings}
\thispagestyle{empty}
\vspace*{-1cm}
{\flushleft\includegraphics[width=3cm,viewport=0 -30 200 -20]{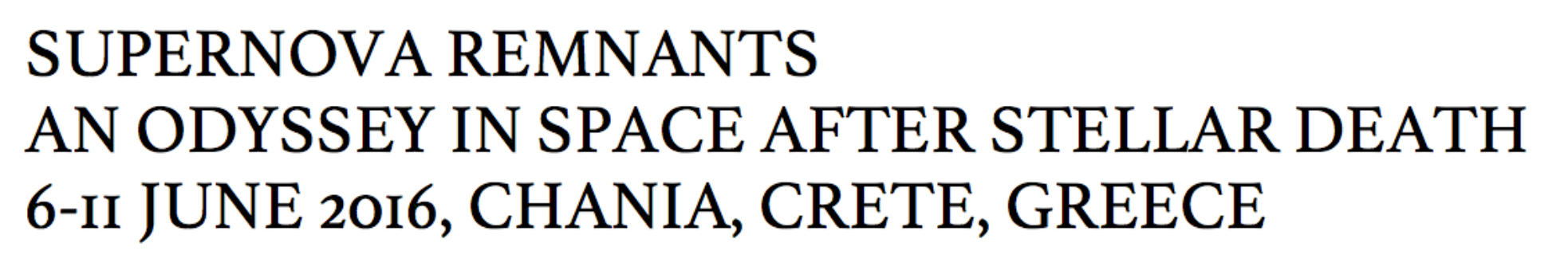}}
\vspace*{0.2cm}

\begin{flushleft}
{\bf {\LARGE
Investigating the X-ray and Gamma-ray Properties of the Galactic Supernova Remnants Kes 69, 3C 396, 3C 400.2
}\\
\vspace{0.3cm}
T\*l\*n Ergin$^1$,
Aytap Sezer$^2$
and Ryo Yamazaki$^3$
}\\
\vspace*{0.2cm}
%
{\footnotesize
$^{1}$
TUBITAK Space Technologies Research Institute, ODTU Campus, 06800, Ankara, Turkey \\
$^{2}$
Harvard-Smithsonian Center for Astrophysics, 60 Garden Street, Cambridge, MA 02138, USA \\
$^{3}$
Department of Physics and Mathematics, Aoyama Gakuin University, 5-10-1, Fuchinobe, Sagamihara 252-5258, Japan \\
}
\end{flushleft}

\markboth{
Investigating the X-ray and Gamma-ray Properties of Kes 69, 3C 396, 3C 400.2
}
{
Ergin et al.
}
\thispagestyle{empty}

\vspace{-1.0cm}
\section{Abstract}
\vspace{-0.3cm}
Kes 69, 3C 396, and 3C 400.2 are mixed-morphology (MM) Galactic supernova remnants (SNRs), where Kes 69 and 3C 396 are interacting with molecular clouds (MCs). Previous X-ray studies showed that the emission from these SNRs is thermal. It has been suggested that MM SNRs interacting with MCs are potential candidates for recombining plasma (RP) in X-rays and hadronic $\gamma$-ray emission. Recently, {\it Chandra} observations revealed signs of RP in 3C 400.2. Our preliminary analyses show that the X-ray emission of NW and SE region of 3C 400.2 arises from recombining plasma. We detected GeV $\gamma$-ray emission from Kes 69 and 3C 396 above 5$\sigma$. 

\vspace{-0.5cm}
\section{Introduction}
\vspace{-0.3cm}
There are several mixed-morphology (MM) Galactic supernova remnants (SNRs) interacting with molecular cloud (MCs) and emitting $\gamma$-ray emission. They originate from a core-collapse (CC) supernova explosion and also show recombining plasma (RP) (e.g., W49B: \citealt{Oz09}, IC 443: \citealt{Ya09}, and 3C 391: \citealt{Er14}). We therefore expect to find several MM SNRs showing RP. Kes 69 (G21.8$-$0.6), 3C 396 (G39.2$-$0.3), and 3C 400.2 (G53.6$-$2.2) are all MM CC SNRs. 
Kes 69 and 3C 396 are interacting with MCs \citep {He09}.  
In this work, we investigate the X-ray and $\gamma$-ray properties of these SNRs using archival {\it Suzaku} and Fermi-LAT data.

\vspace{-0.5cm}
\section{X-ray Observations and Analysis}
\vspace{-0.2cm}
These SNRs observed with X-ray Imaging Spectrometer \citep[XIS;][]{Ko07} onboard {\it Suzaku}. Their observation logs are listed in Table 1. The data reduction and analysis were carried out with {\sc HEAsoft} package version 6.16 and {\sc xspec} version 12.9.0 \citep {Ar96} with AtomDB 3.0.3 \citep {Sm01, Fo12}.

The background spectra were extracted from the source-free region in the same field of view. The background emission consists of the instrumental non X-ray background (NXB), cosmic X-ray background (CXB) and Galactic ridge X-ray emission (GRXE) for these SNRs. We estimated the NXB spectra using the tool {\sc xisnxbgen} \citep {Ta08}. 

The source spectra were extracted from circular and ellipses regions for these SNRs. For spectral fitting, we used a non-equilibrium ionization (NEI) recombining collisional plasma model with variable abundances (VRNEI in {\sc xspec}). We used the absorption TBABS model in {\sc xspec} \citep {Wi00} and abundances of \citet {Wi00}.  
\begin{table*}
 \begin{minipage}{170mm}
  \begin{center}
  \caption{Log of the {\it Suzaku} observations}
   \vspace{0.2cm}
\begin{tabular}{@{}lcccc@{}}
  \hline
\hline
SNRs  &  Obs.ID  & Obs.Date &  Exposure (ks)     \\
\hline
Kes 69  & 509037010 & 2014-09-27  &  77.4  \\
3C 396&  509038010 & 2014-04-26  & 82.8  \\
3C 400.2 NW &509068010 &2014-04-23&  21.5  \\
3C 400.2 SW & 509069010 & 2014-04-14 &   24.2  \\
3C 400.2 SE &509070010 & 2014-04-23 & 24.9  \\
3C 400.2 NE & 509071010 & 2014-04-23 &  20.2 \\
  \hline
\end{tabular}
\label{table_1}
\end{center}
\end{minipage}
\end{table*}
\begin{table*}
 \begin{minipage}{150mm}
  \begin{center}
   \caption{Log of the {\it Fermi}-LAT observations}
 \vspace{0.2cm}
\begin{tabular}{@{}lccccc@{}}
  \hline
\hline
SNR Name       &  Obs. Start Date  & Obs. End Date   & ROI R.A.(J2000) & ROI Decl.(J2000)\\
\hline
Kes 69             & 4 August 2008      & 15 May 2016      & 18$^h$ 32$^m$ 45$^s\!$.12  & $-$10$^{\circ}$ 08$'$ 00$\farcs$12\\
3C 396             & 4 August 2008      &  14 May 2016     & 19$^h$ 04$^m$ 08$^s\!$.12  & $+$05$^{\circ}$ 28$'$ 00$\farcs$12\\
3C 400.2          & 4 August 2008      &  6 January 2015 & 19$^h$ 38$^m$ 50$^s\!$.00  & $+$17$^{\circ}$ 14$'$ 00$\farcs$12\\
  \hline
\end{tabular}
\label{table_5}
\end{center}
\end{minipage}
\end{table*}

\vspace{-0.6cm}
\section{$\gamma$-ray Analysis}
\vspace{-0.2cm}
To search for a $\gamma$-ray counterpart of the SNRs Kes 69, 3C 396, and 3C 400.2 in the GeV energy range, we analyzed the $\gamma$-ray data of Large Area Telescope (LAT) on board {\it Fermi} Gamma Ray Space Telescope ({\it Fermi}) for the time periods as given in Table \ref{table_5}. 

Using \texttt{gtselect} of Fermi Science Tools, we selected the {\it Fermi}-LAT Pass 8 'Source' class and front$+$back type events coming from zenith angles smaller than 90$^{\circ}$ and from a circular region of interest (ROI) with a radius of 20$^{\circ}$ centered at the corresponding SNR position as shown in Table \ref{table_5}. 

We applied the maximum likelihood fitting method on the spatially and spectrally binned data within a square region of 28$^{\circ}$ $\times$ 28$^{\circ}$ and used the instrument response function version P8R2$_{-}\!\!$SOURCE$_{-}\!\!$V6. While modeling the analysis region, the diffuse Galactic emission (\emph{gll$_{-}$iem$_{-}$v6.fits}) and the isotropic component (\emph{iso$_{-}$P8R2$_{-}\!\!$SOURCE$_{-}\!\!$V6$_{-}$v06.txt}) were taken into account, where the normalization of the isotropic component and the normalization of the galactic diffuse background were set free during the analysis. The model also included all point-like and extended sources from the 3rd {\it Fermi}-LAT Source Catalog \citep{Ac15} located within a distance of 20$^{\circ}$ from the ROI center. 
We fixed the parameters of all sources that are located at distances greater than 7$^{\circ}$ from the ROI center. The normalization of all variable background sources with variability values higher than $\sim$72 and all parameters of sources with TS values above 120 were set free. 

\begin{figure*}
\centering \vspace*{1pt}
\includegraphics[width=0.30\textwidth]{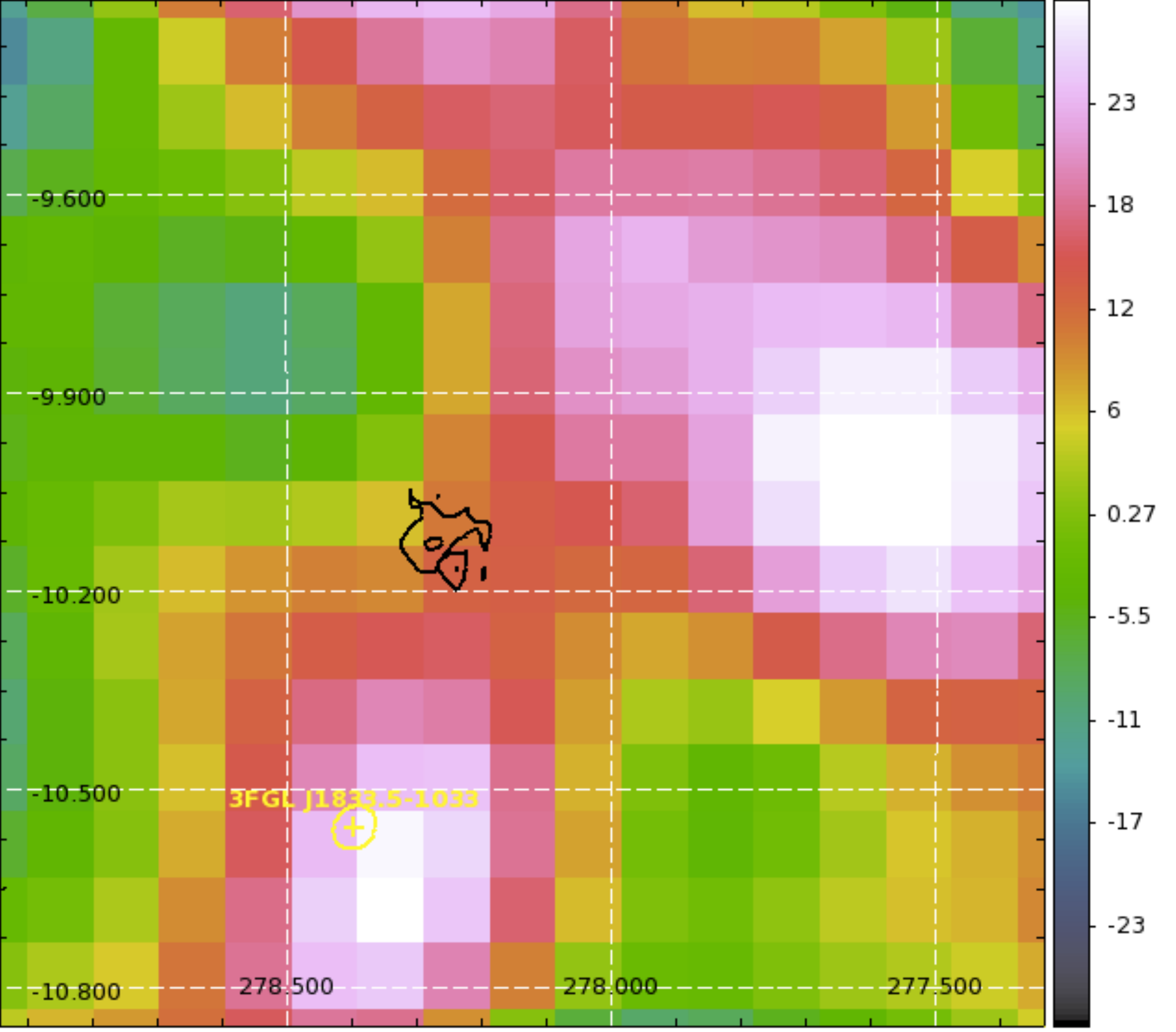}
\includegraphics[width=0.30\textwidth]{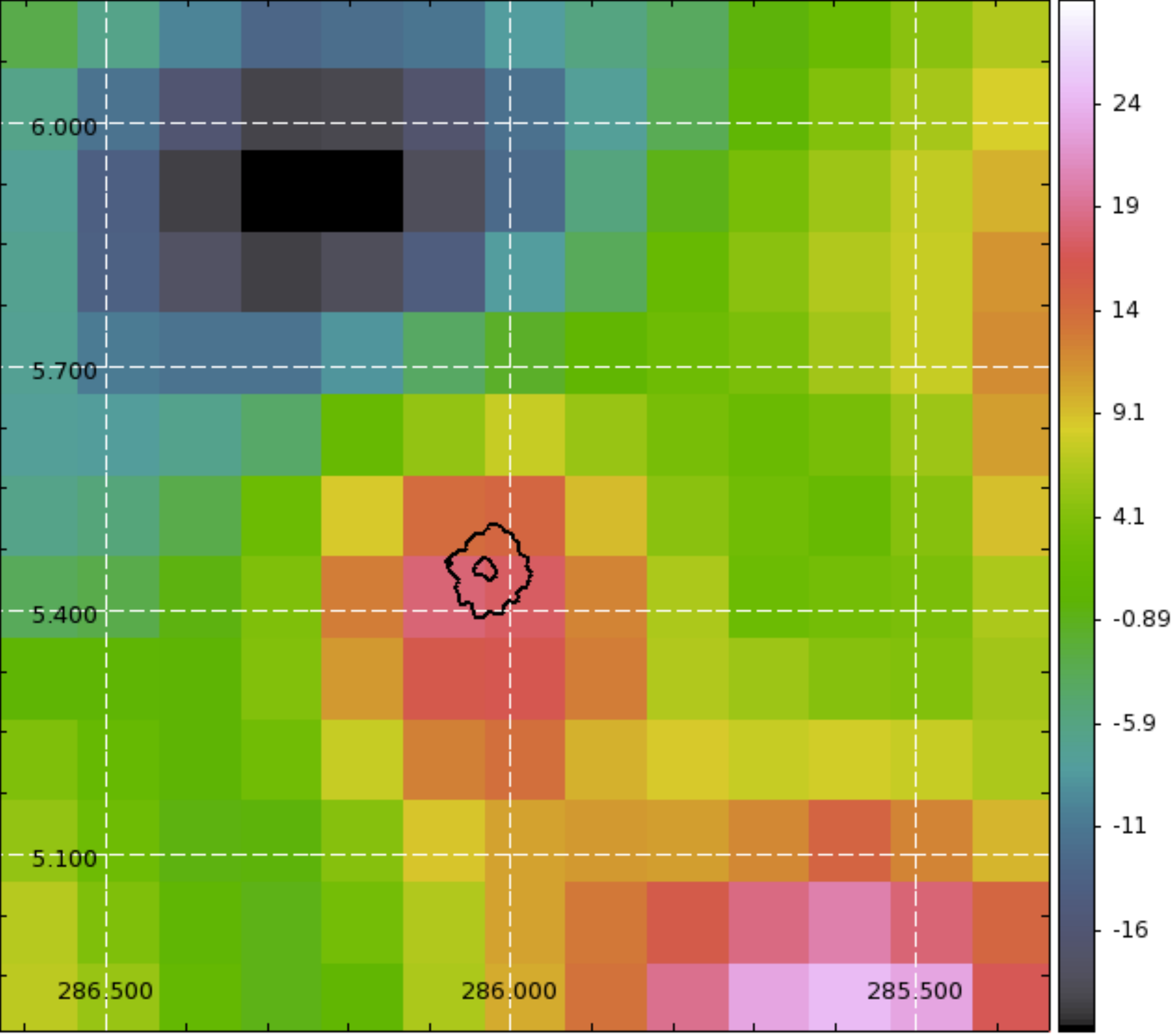}
\includegraphics[width=0.304\textwidth]{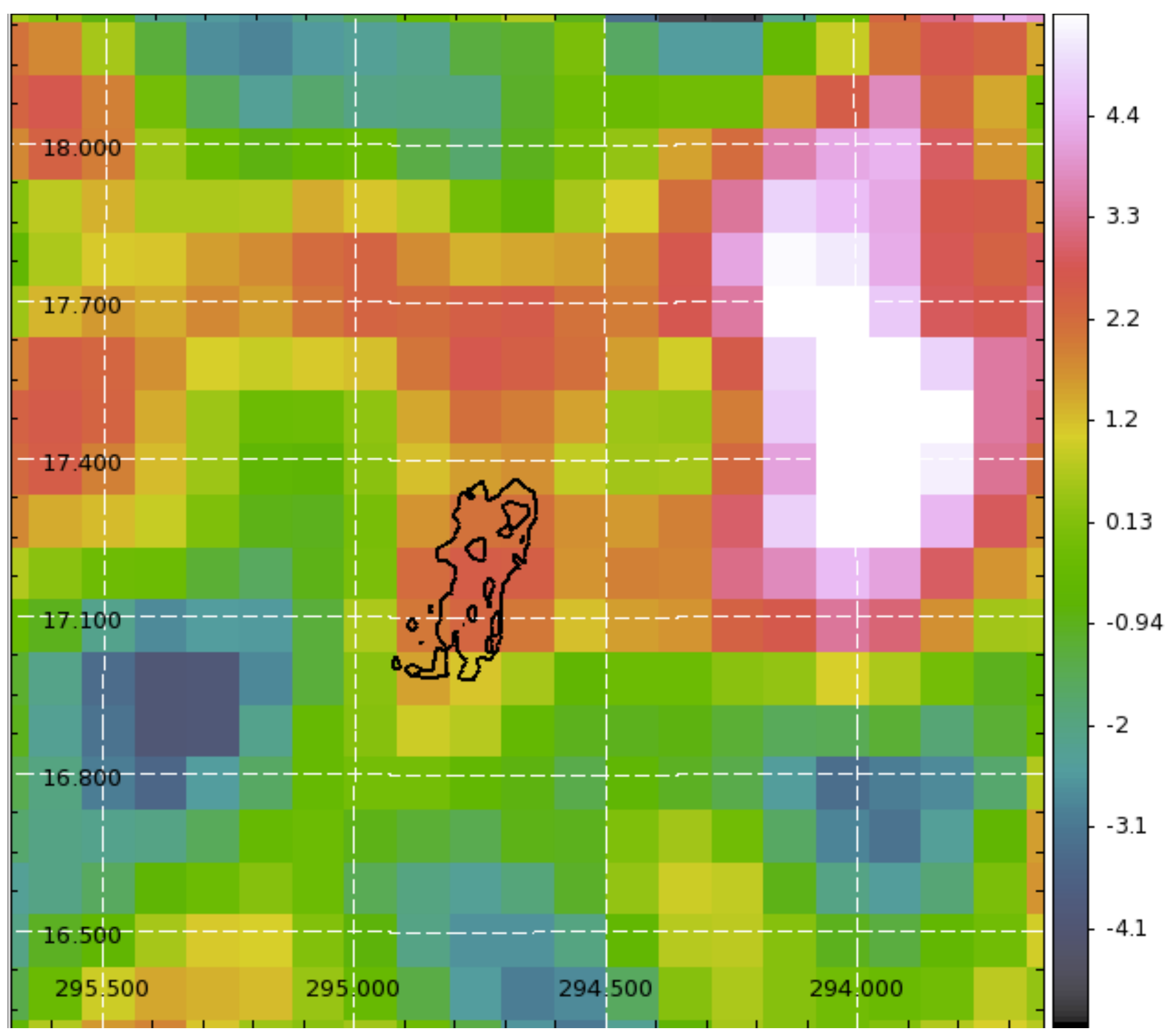}
\caption{ {\footnotesize {\it Left Panel}: The gamma-ray residual map of Kes 69 (left), 3C 396 (center), and 3C400.2 (right). The residual maps are smoothed with a 3-sigma Gaussian kernel. The black contours are from {\it Suzaku} with 8.0, 9.0,10.0 counts for Kes 69, 0.0, 9.3 and 18.6 counts for 3C 396 and 3.4, 4.2, 5.0 for 3C400.2. }}
\label{figure_3}
\end{figure*}

\vspace{-0.5cm}
\section{Preliminary Results and Discussion}
\vspace{-0.3cm}

In this work, we present the {\it Suzaku} and {\it Fermi}-LAT preliminary analysis results of the MM SNRs Kes 69, 3C 396, and 3C 400.2. 

We found that the overionized plasma in SE and NW regions of 3C 400.2. The X-ray emission of SW and NE regions is well represented by a two thermal plasma model; a high-temperature in NEI and a low-temperature in collisional ionization equilibrium (CIE) condition.
Using Chandra data, \citet {Br15} found overionizing plasma in all parts of this remnant. 3C 400.2 showed no significant $\gamma$-ray emission, when fitted as a point-like source at the radio-location with a simple power-law type spectrum. The nearby region of 3C 400.2 contains a spot of brighter $\gamma$-ray excess (Figure \ref{figure_3} right panel), which needs to be further analyzed using more data.  

The X-ray spectrum of 3C 396 is described by two component thermal model with a column density $N_{\rm H}$ $\sim$ 6.1$\times$10$^{22}$ cm$^{-2}$. The $\gamma$-ray analysis revealed significant $\gamma$-ray emission at the radio-SNR location of 3C 396 with a significance of $\sim$25$\sigma$ assuming 3C 396 to be a point-like source with a power-law type energy spectrum. Since the analysis region contains areas of excess $\gamma$-ray residuals (Figure \ref{figure_3} center panel), it requires further investigation. 
 
We found that the X-ray emission from Kes 69 is well fitted by a combination of NEI and CIE model with absorbing column density of $N_{\rm H}$ $\sim$ 2.8$\times$10$^{22}$ cm$^{-2}$. Kes 69 was detected $\sim$16$\sigma$ by modeling the SNR as a point-like $\gamma$-ray source having a simple power-law type spectrum. There is a 3rd {\it Fermi}-LAT catalog source (3FGL J1833.5$-$1033) about a distance of 0.47$^{\circ}$ away from Kes 69 that exhibits an excess of $\gamma$ rays (Figure \ref{figure_3} left panel). Also there is another $\gamma$-ray excess region very close to Kes 69, where no counterparts exist. So, the analysis needs to be refined by taking the excess $\gamma$-ray emission in the neighborhood of Kes 69 into account. This will help us to better characterize the $\gamma$-ray emission originating from Kes 69. 

\small
\vspace{-0.5cm}
\section*{Acknowledgments}   
\vspace{-0.3cm}
TE thanks to the support by the Young Scientists Award Program (BAGEP-2015). AS is supported by the Scientific and Technological Research Council of Turkey (T\"{U}B\.{I}TAK) through the B\.{I}DEB-2219 fellowship program. RY is supported in part by grant-in-aid from the Ministry of Education, Culture, Sports, Science, and Technology (MEXT) of Japan, No. 15K05088.


\end{document}